\documentclass[a4paper,UKenglish,cleveref,autoref,thm-restate]{oasics-v2021}

\pdfoutput=1 

\title{Generating Software for Well-Understood Domains}

\author{Jacques Carette}{Department of Computing and Software, McMaster University, 1280 Main Street West, Hamilton, Ontario, L8S 4L8, Canada \and \url{https://www.cas.mcmaster.ca/~carette/} }{carette@mcmaster.ca}{https://orcid.org/0000-0001-8993-9804}{}
\author{Spencer Smith}{Department of Computing and Software, McMaster University, 1280 Main Street West, Hamilton, Ontario, L8S 4L8, Canada \and \url{https://www.cas.mcmaster.ca/~smiths/} }{smiths@mcmaster.ca}{https://orcid.org/0000-0002-0760-0987}{}
\author{Jason Balaci}{Department of Computing and Software, McMaster University, 1280 Main Street West, Hamilton, Ontario, L8S 4L8, Canada}{balacij@mcmaster.ca}{}{}

\authorrunning{J. Carette, S. Smith, and J. Balaci}

\Copyright{Jacques Carette, Spencer Smith, and Jason Balaci}



\category{Research} 

\relatedversion{}

\nolinenumbers 

\EventEditors{Ralf L\"{a}mmel, Peter D. Mosses, and Friedrich Steimann}
\EventNoEds{3}
\EventLongTitle{Eelco Visser Commemorative Symposium (EVCS 2023)}
\EventShortTitle{EVCS 2023}
\EventAcronym{EVCS}
\EventYear{2023}
\EventDate{April 5, 2023}
\EventLocation{Delft, The Netherlands}
\EventLogo{}
\SeriesVolume{109}
\ArticleNo{1}

\newcommand{\CC}{C\nolinebreak\hspace{-.05em}\raisebox{.4ex}{\small\bf +}\nolinebreak\hspace{-.10em}\raisebox{.4ex}{\small\bf +}}

\usepackage{dashbox}

\begin{document}

\maketitle

\begin{abstract}
    Current software development is often quite code-centric and aimed at
    short-term deliverables, due to various contextual forces (such as the
    need for new revenue streams from many individual buyers). We're interested
    in software where different forces drive the development. \textbf{Well
    understood domains} and \textbf{long-lived software} provide one such
    context.

    A crucial observation is that software artifacts that are currently
    handwritten contain considerable duplication.  By using domain-specific
    languages and generative techniques, we can capture the contents of many of
    the artifacts of such software.  Assuming an appropriate codification of
    domain knowledge, we find that the resulting de-duplicated sources are
    shorter and closer to the domain.  Our prototype, Drasil, indicates
    improvements to traceability and change management. We're also hopeful
    that this could lead to long-term productivity improvements for software
    where these forces are at play.
\end{abstract}

\section{The Context}
\label{sec:the-context}

Not all software is the same. In fact, there is enough variation in the
context in which developers create various software products to warrant exploring and
using different processes, depending on the forces~\cite{alexander1977pattern}
at play. Here we explore two such forces: ``well understood'' and ``long lived''.

We need to be precise about what we mean by ``well understood'' and ``long lived'',
and thus will start by defining these terms. The definition of ``well understood''
has three components, which will be successively refined throughout this paper.
As the codification of what is understood tends to be reliably communicated via
\emph{documentation}, we review that as well, before discussing which aspects of
software products, aka software artifacts, we are considering. Then we give
concrete examples of software domains where these ideas come together.

This work can be seen as resurrecting many of the goals of projects like
Draco~\cite{neighbors1984draco,neighbors1989draco}, 
DMS~\cite{baxter1992design,baxter2002dms}, 
GLisp~\cite{novak1983glisp,novak1994generating} and 
SpecWare~\cite{srinivas95specware,smith1999mechanizing}. 
All are
variants on \emph{automatic programming}. Unlike many of these projects, our
primary aim is not automation, but rather a combination of raising the level
of abstraction and knowledge capture. And indeed \emph{domain engineering},
as coined by Draco's creator James Neighbors, is core to our work. Where we
differ is our emphasis on all software artifacts, not just code, and being able to leverage
several decades of theoretical and technological advances. See~\ref{sec:relwork}
for more details on these and other related projects and ideas.

\subsection{``Well-understood'' software?}
\label{subsec:well-understood}

\begin{definition}
\label{defn:well-understood}
A software domain is \textbf{well understood} if
\begin{enumerate}
    \item its Domain Knowledge (DK)~\cite{bjorner2021domaineng} is codified,
    \item the computational interpretation of the DK is clear, and
    \item writing code to perform said computations is well understood.
\end{enumerate}
\end{definition}

By \emph{codified}, we mean that the knowledge exists in standard form in a
variety of textbooks. For example, many engineering domains use ordinary
differential equations as models, the quantities of interest are known, given
standard names and standard units. In other words, standard vocabulary has been
established over time and the body of knowledge is uncontroversial.

We can refine these high level ideas, using the same numbering, although the
refinement should be understood more holistically.
\begin{enumerate}
\item Models in the DK \emph{can be} written formally.
\item Models in the DK \emph{can be} turned into functional relations by
 existing mathematical steps.
\item Turning these functional relations into code is an understood
 transformation.
\end{enumerate}
Most importantly, the last two parts deeply involve \emph{choices}: What
quantities are considered inputs, outputs and parameters to make the model
functional? What programming language?  What software architecture,
data-structures, algorithms, etc.?

In other words, \emph{well understood} does not imply \emph{choice free}.
Writing a small script to move files could just as easily be done in Bash,
Python or Haskell. In all cases, assuming fluency, the author's job is
straightforward because the domain is well understood.

\subsection{Long-lived software?}
\label{subsec:long-lived-software}

For us, long-lived software~\cite{SPL-long-lived} is software that is expected
to be in continuous use and evolution for \(20\) or more years. The main
characteristic of such software is the \emph{expected turnover} of key staff.
This means that all tacit knowledge about the software will be lost over time if
it is not captured. Relying on oral tradition and word-of-mouth becomes untenable
for the viability of long-lived projects, and documentation becomes a core asset of
value comparable, if not greater than, the actual code.

\subsection{Documentation}
\label{subsec:documentation}

Well understood also applies to \textbf{documentation} aimed
at humans~\cite{parnas2011precise}. The same three items of
Definition~\ref{defn:well-understood} can be retargeted to apply to
documentation as follows: (1) The meaning of the models is understood at a
human-pedagogical level, i.e., it is explainable.  (2)~Combining models is
explainable. Thus, the act of combining models must simultaneously operate on
mathematical representations and on explanations. This requires that English
descriptions also be captured in the same manner as the formal-mathematical
knowledge.  (3)~The refinement steps that are performed due to making
software-oriented decisions should be captured with a similar mechanism, and
also include English explanations.

Coupling together well-understood domains and its documentation, we
obtain what we call triform theories.
\begin{definition}
\label{defn:triform-theories}
\textbf{Triform theories}, as a nod to \emph{biform
theories}~\cite{Farmer2007}, are the coupling of three concepts:
\begin{enumerate}
  \item an axiomatic description,
  \item a computational description, and
  \item an English description.
\end{enumerate}
\end{definition}
\noindent These will form the heart of our approach to domain engineering.

\subsection{Software artifacts}
\label{subsec:software-artifacts}

Software currently consists of a whole host of artifacts: requirements,
specifications, user manual, unit tests, system tests, usability tests, build
scripts, READMEs, license documents, process documents, as well as code.
Our aim is to understand the ``information content'' of each artifact
sufficiently to construct a Domain Specific Language (DSL) that describes how
to weave information gleaned from triform theories with instance specific
choices that are sufficient to enable the generation of each artifact.

\subsection{Instances}
\label{subsec:examples-of-context}

When are these well understood and long lived conditions fulfilled? One example
is \emph{research software} in science and engineering. While the results of
running various simulations is entirely new, the underlying models and how to
simulate them are indeed well known. One particularly long-lived example is
embedded software for space probes (like Pioneer 10). Another would be ocean
models, such as NEMO~\cite{madec_gurvan_2022_6334656}, which can trace its
origins to OPA 8~\cite{Madec1998}. In fact, most subdomains of science and
engineering harbour their own niche long-lived software. The domain of
control systems is particularly rich in examples.

Note that some better-known long-lived software, such as financial
systems, do not fall within our purview, as the implemented  business processes
are not \emph{well understood}. This is why most large rewrites of financial 
systems fail as the functionality actually implemented in the source system is
not documented.

\begin{figure*}[t!]
  \centering
  \includegraphics[width=\linewidth]{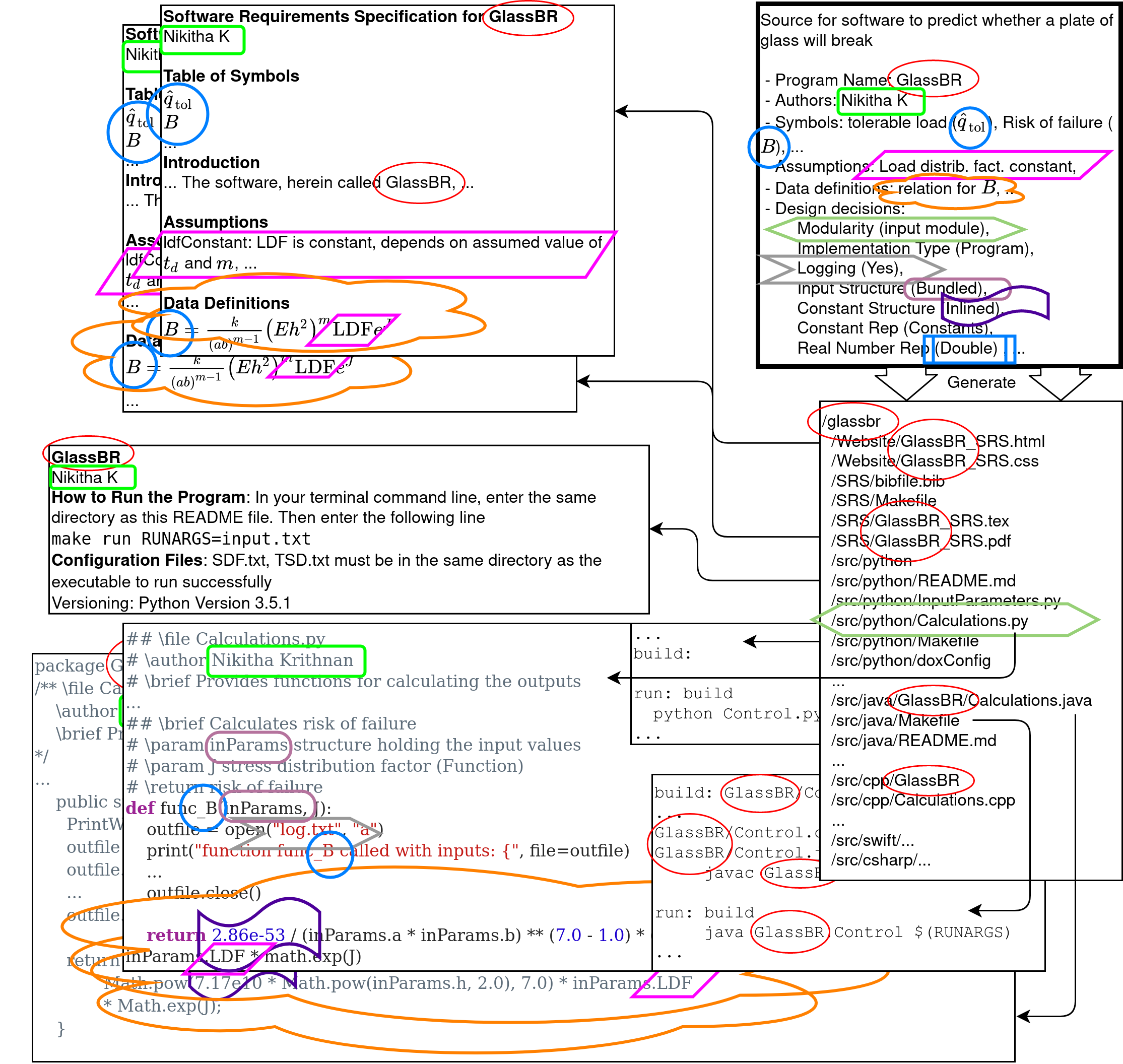}
  \caption{Colors and shapes mapping from captured domain knowledge to generated
  artifacts. Red oval: program name; green rectangle: author name; blue circle:
  symbols for real-valued variables; pink lozenge: assumptions used (load 
  distribution factor is constant);
  orange cloud: mathematical definition of $B$; gray arrow: option to turn on
  logging; light brown rounded-corners rectangle: how inputs should be ``bundled'';
  purple wavy rectangle: how constants are handled (here: inlined).}
  \label{Fig_DrasilAndChange}
\end{figure*}

\section{An Example}
\label{sec:example}

We have built infrastructure, which we also call Drasil%
\footnote{\url{https://github.com/JacquesCarette/Drasil}} 
to carry out these ideas. It consists of 60KLoc of Haskell
implementing a series of interacting Domain Specific Languages (DSLs) for
knowledge encodings, mathematical expressions, theories, English fragments,
code generation and document generation.  

Drasil consists of a ``database'' of information about certain domains of
interest (parts of physics, some mathematics, software engineering,
documentation and computing), encoded in a home-grown ontology somewhat
resembling the work of Novak~\cite{kook1991representation,novak1994generating}.
It has also a number of internal DSLs for representing the contents of
various kinds of documents (specifications, programs, simple build scripts,
READMEs), coupled with DSLs for rendering in various formats (\LaTeX, HTML
and plain text, used for specifications, and GOOL~\cite{GOOLPEPM} for
representing Object-Oriented programs that can be rendered in six different
languages). There are also DSLs for triform theories, mathematical expressions,
English fragments, citations, authors, some software choices, and so on. Lastly
another DSL weaves together all the previous information to render the
artifacts that together make a piece of software. A longer article describing
all of these in detail is being written.

We provide a bit of the overall flavour of Drasil through what we hope to be
an illustrative example.

\subsection{GlassBR}
We will focus on information capture and the artifacts we can generate. For
concreteness, we'll use a single example from our suite: GlassBR, used to assess
the risk for glass facades subject to blast loading. The requirements are based
on an American Standard Test Method (ASTM) standard \cite{ASTM2009, ASTM2015,
BeasonEtAl1998}. GlassBR was originally a Visual Basic code/spreadsheet
created by colleagues in a civil engineering research lab.  We added their
domain knowledge to our framework, along with recipes to generate relevant
artifacts.  Not only can we generate code for the necessary calculations (in
\CC, C\#, Java, Python and Swift), we also generated documentation that was absent in the
original (Software Requirements Specification, doxygen, README.md and a
Makefile). Moreover, our implementation is actually a family of implementations,
since some design decisions are explicitly exposed as changeable variabilities,
as described below.

Figure~\ref{Fig_DrasilAndChange} illustrates the transformation of captured
domain knowledge. Reading this figure starts from the upper right box. Each
piece of information has its own shape and colour (orange cloud,
pink lozenge, etc). It should be immediately clear that all pieces of
information reappear in multiple places in the generated artifacts. For example,
the name of the software (GlassBR) ends up appearing more than 80 times in the
generated artifacts (in the folder structure, requirements, README, Makefile and
source code). Changing this name would traditionally be extremely difficult; we
can achieve this by modifying a single place, and regenerating.

The first box shows the directory structure of the currently generated
artifacts; continuing clockwise, we see examples of Makefiles for the Java and
Python versions, parts of the fully documented, generated code for the main
computation in those languages, user instructions for running the code, and the
processed \LaTeX{} for the requirements.

The name GlassBR is probably the simplest example of what we mean by
\emph{duplication}: here, the concept ``program name'' is internally defined, and
its \emph{value} is used throughout. 

In general, we capture more complex knowledge. An example is the assumption
that the ``Load Distribution Factor'' (LDF) is constant (pink lozenge). If this
needs to be modified to instead be an input, the generated software will now
have LDF as an input variable.  

We also capture design decisions, such as
whether to log all calculations, whether to inline constants rather than show
them symbolically, etc. These different pieces of knowledge can also be reused
in different projects.

\subsection{The Steps}
We describe an ``idealized process'' that we could have used to produce GlassBR,
following Parnas' idea of faking a rational design process \cite{Parnas1986}.

\subparagraph*{Understand the Program's Task} Compute the probability that a
particular pane of (special) glass will break if an explosive is detonated at a
given distance.  This could be in the context of the glass facade for a
building.

\subparagraph*{Is it well understood?} The details are extensively documented
in~\cite{ASTM2009, ASTM2015, BeasonEtAl1998}.

\subparagraph*{Record Base Domain Knowledge}
A recurring idea is the different types of \texttt{Glass}:
\begin{center}
  \begin{tabular}{|l|l|l|l|}
    \hline
    \textbf{Concept} & \textbf{Term (Name)} & \textbf{Abbrev.} & \textbf{Domain} \\ \hline
    \texttt{fullyT} & Fully Tempered & FT & \texttt{[Glass]} \\ \hline
    \texttt{heatS} & Heat Strengthened & HS & \texttt{[Glass]} \\ \hline
    \texttt{iGlass} & Insulating Glass & IG & \texttt{[Glass]} \\ \hline
    \texttt{lGlass} & Laminated Glass & LG & \texttt{[Glass]} \\ \hline
    \texttt{glassTypeFac} & Glass Type Factor & GTF & \texttt{[Glass]} \\ \hline
  \end{tabular}
\end{center}

\noindent The ``Risk of Failure'' is definable:
\begin{center}
  \begin{tabular}{|l|l|}
    \hline
    \textbf{Label} & Risk of Failure \\ \hline
    \textbf{Symbol} & $B$ \\ \hline
    \textbf{Units} & Unitless \\ \hline
    \textbf{Equation} & \(B = \frac{k}{(ab)^{m-1}}(Eh^2)^m\mathit{LDF}e^J\) \\ \hline
    \textbf{Description} & \vbox{
      \hbox{\strut \(B\) is the Risk of Failure (Unitless)}
      \hbox{\strut \(k\) is the surface flaw parameter (\(\frac{m^{12}}{N^7}\))}
      \hbox{\strut \(a\) \& \(b\) are the plate length \& width (\textit{m})}
      \hbox{\strut $m$ is the surface flaw parameter ($\frac{m^{12}}{N^7}$)}
      \hbox{\strut $E$ is the modulus of elasticity of glass (\textit{Pa})}
      \hbox{\strut $h$ is the minimum thickness (\textit{m})}
      \hbox{\strut $LDF$ is the load duration factor (Unitless)}
      \hbox{\strut $J$ is the stress distribution factor (Unitless)}
    } \\ \hline
    \textbf{Source} & \cite{ASTM2009}, \cite{BeasonEtAl1998} \\ \hline 
  \end{tabular}
\end{center}
Some concepts, such as those of \emph{explosion}, \emph{glass slab}, and
\emph{degree} do not need to be defined mathematically --- an English description
is sufficient.

The descriptions in GlassBR are produced using an experimental language using
specialized markup for describing relations between knowledge. For example, the
goal of GlassBR (``Predict-Glass-Withstands-Explosion'') is to ``Analyze and
predict whether the \textit{glass slab} under consideration will be able to
withstand the \textbf{explosion} of a certain \textbf{degree} that is
calculated based on \textit{user input}.'', where italicized names are ``named
ideas'', and bold faced names are ``concept chunks'' (named ideas with a domain
of related ideas). We call this goal a ``concept instance'' (a concept chunk
applied in some way). This language lets us perform various static analyses on
our artifacts.

This doesn't build a complete ontology of concepts, as we have not found that
to be necessary to generate our artifacts. In other words, we define a
\emph{good enough} fraction of the domain ontology.

The most important results of this phase are descriptions of \emph{theories}
that link all the important concepts together. For example, the description
of all the elements that comprise Newton's Law $F = m a$,
i.e., what is a force, a mass, acceleration (and how is it related to
velocity, position and time), what are the units involved, etc.

\subparagraph*{Define the characteristics of a good solution}
For example, one of our outputs is a probability, which means that the output
should be checked to be between \(0\) and \(1\). This can result
in assertion code in the end program, or tests, or both. We do not yet
support full \emph{properties}~\cite{claessen2000quickcheck}, but that is
indeed the logical next step.

\subparagraph*{Record basic examples}
For the purposes of testing, it is always good to have very simple examples,
especially ones where the correct answer is known a priori, even though the simple
examples are considered ``toy problems''. They provide a useful extra check
that the narrative is coherent with our expectations.

\subparagraph*{Specialization of theories}
In general, the theories involved will be much more general than what is
needed in any given example. For example, Newton's Laws are encoded in their
vector form in $n$ dimensions, and thus specialization is necessary.

In the GlassBR example, the thickness parameter
is not free to vary, but must take one of a specific set of values. The
rationale for this specialization is that manufacturers have standardized the
glass thickness they will provide. (This rationale is also something we capture.)

Most research software examples involve significant specialization, such as
converting partial differential equations to ordinary, elimination of
variables, use of closed forms instead of implicit equations, and so on.
Often specialization, driven by underlying assumptions, enable further
specializations, so that this step is really one of \emph{iterative refinement}.

\subparagraph*{Create a coherent narrative}
Given the outputs we wish to produce, such as the probability that a glass
slab will withstand an explosion, we need to ensure that we can go from all
the given inputs to the desired output, by stringing together various
definitions given in a computational manner. In other words, we need to
ensure that there exists a deterministic path from the inputs we are given,
through the equations we have, to all of the outputs we've declared we are
interested in.

For this example, the computations are all quite simple. The reasons why
GlassBR's few lines of code are correct is what is interesting, and that
description, which is part of the specification, spans a few pages. That
information is generic about the properties of glass and the effects of
explosions, and could be reused in other applications.
In general, research software tends to involve
solving ordinary differential equations, computing a solution to
an optimization problem, etc. Again the reason why a particular differential
equation is an adequate model of the problem at hand is part of the
important information we wish to capture.

In other words, the main narrative of a piece of software is its high level
design. What are the techniques used to derive the outputs from the inputs,
obtained in a reasoned manner by successive refinements of the underlying
theory into an actual program. While the code itself is akin to the actual
words used to tell a story, the narrative is the story, its themes
and topics, and so on.

\subparagraph*{Make code level choices}
From a \emph{deterministic model of the solution}, it should be
possible to output code in a programming language. To get there, we still need
to make a series of choices.

Drasil lets you choose output programming language(s) (see
\Cref{fig:CodeChoices}), but also how ``modular'' the generated code is, whether
we want programs or libraries, the level of logging and comments, etc. Here we
show the actual code we use for this, as it is reasonably direct.

\begin{figure}[htb]
\begin{lstlisting}
code :: CodeSpec
code = codeSpec fullSI choices allMods

choices :: Choices
choices = defaultChoices {
 lang = [Python, Cpp, CSharp, Java, Swift], 
 modularity = Modular Separated,
 impType = Program, logFile = "log.txt", 
 logging = [LogVar, LogFunc],
 comments = [CommentFunc, CommentClass, CommentMod], 
 doxVerbosity = Quiet,
 dates = Hide, 
 onSfwrConstraint = Exception, onPhysConstraint = Exception,
 inputStructure = Bundled, 
 constStructure = Inline, constRepr = Const
}
\end{lstlisting}
  \caption{Code level choices for GlassBR (corresponds to much of the material
  in the bold box in top right of \Cref{Fig_DrasilAndChange}).}
  \label{fig:CodeChoices}
\end{figure}

\subparagraph*{Create recipes to generate artifacts}
The information collected in the previous steps form the core of the various
software artifacts normally written. To perform this assembly, we write
programs that we dub \emph{recipes}. We have DSLs for creating 
requirements specifications~\cite{SmithEtAl2007},
code~\cite{GOOLPEPM}, dependency diagrams, Makefiles, READMEs and a log of all choices
used.


Finally, we can put the contents created via the previous steps together and
\textbf{generate everything}, here the SRS, the code, traceability graphs and
a log of the design choices.

\section{An Idealized Process}
\label{sec:idealized-process}

The above used what we refer to as an \emph{idealized process} for
the development of software that is both well-understood and
long-lived. Drasil is meant to facilitate this process. In other words,
as we ourselves better understand this idealized process, we modify Drasil
to follow suit, rather than the other way around. We can summarize the
process as follows.

\begin{enumerate}
\item\label{it:problem} Have a task to achieve where \emph{software} can
play a central part in the solution.
\item\label{it:understood} Verify that the underlying problem domain is \emph{well
understood}.
\item\label{it:probdesc} Describe the problem:
  \begin{enumerate}
  \item Find the base knowledge (theories) in the pre-existing library or,
    failing that, write it if it does not yet exist, for instance the
    naturally occurring known quantities and associated constraints.
  \item Describe the characteristics of a good solution.
  \item Come up with basic examples (to test correctness, intuitions, etc).
  \end{enumerate}
\item\label{it:refine} Describe, by successive refinement transformations,
how the problem description can be turned into a deterministic\footnote{A current
meta-design choice.} input-output process.
  \begin{enumerate}
  \item Some refinements will involve \emph{specialization} (e.g., from
    \(n\)-dimensional to \(2\)-dimensional, assuming no friction, etc).  These
    \emph{choices} and their \emph{rationale} need to be documented, as a
    crucial part of the solution.  Whether these choices are (un)likely to
    change in the future should also be recorded.
  \item Choices tend to be dependent, and thus (partially) ordered.
   \emph{Decisions} frequently enable or reveal downstream choices.
  \end{enumerate}
\item\label{it:narrative} Assemble the ingredients into a coherent narrative.
\item\label{it:tocode} Describe how the process from step \ref{it:narrative}
can be turned into code. Many choices can occur here as well.
\item\label{it:recipe} Turn the steps (i.e.,\ from items~\ref{it:refine} and
\ref{it:tocode}) into a \emph{recipe}, aka program, that weaves together all
the information into a variety of artifacts (documentation, code, build
scripts, test cases, etc). These can be read, or executed, or \ldots\, as
appropriate.
\end{enumerate}

The fundamental reason for focusing on \emph{well-understood} software is to
make the various steps in this process feasible. Another enabler is a
\emph{suitable} knowledge encoding. Rather than define this a priori, we have
used a \emph{bottom up} process to capture ``good enough'' ontologies to get the
job done.

What is missing in the above description is an explicit \emph{information
architecture} of each of the necessary artifact. In other words, what
information is necessary to enable the mechanized generation of each artifact?
It turns out that many of them are quite straightforward.

Note that in many software projects, steps~\ref{it:problem}
and~\ref{it:probdesc} are skipped; this is part of the \textbf{tacit knowledge}
of a lot of software.  Our process requires that we make this knowledge
explicit, a fundamental step in \emph{Knowledge Management}~\cite{Dalkir2011}.


\section{Related Work}\label{sec:relwork}

As one referee put it, our work is ``old fashioned'', in that its goals
indeed repeat that of work from decades ago. What we see, however, is
that only a small slice of those goals has made it into today's
practices. Thus, we think it is worthwhile to revisit these goals.

Draco~\cite{neighbors1984draco,neighbors1989draco}, one of the oldest
such projects, is also one of the closest to Drasil, goal-wise. Many of
its ideas have percolated through the literature on DSLs and product
lines (see the wonderful textbook~\cite{apel2016feature} for a modern
exposition), and we used them without a proper appreciation for their
source. While DMS~\cite{baxter1992design,baxter2002dms}, and
GLisp~\cite{novak1983glisp,DBLP:journals/tse/Novak95a} are different in their
aims, they also contain some of the same ideas (knowledge capture,
using various ontologies, refinement and reuse). The examples that
drove GLisp (see~\cite{novak1994generating,kook1991representation})
are very close to many of ours, namely applications of physics.
SpecWare~\cite{srinivas95specware,smith1999mechanizing} goes much
further along the formality spectrum, incorporating correctness proofs
as part of its refinement process, as well as being based on a solid
theoretical framework.

Unfortunately a lot of the above work was very code-centric. None of
the systems generate all software artifacts, especially not documentation.
Furthermore, quite a lot of the work uses fairly coarse-grained
chunks of information (``components'') which, in our experience, reduces the
potential for reuse. Many of the projects were extremely ambitious regarding
what aspects of software production they aimed to automate. We do not
aim to automate any activity that can reasonably be called \emph{design}.

We have borrowed ideas from product families~\cite{parnas1976families}
and software product lines~\cite{apel2016feature}.  We have also
borrowed ideas from literate programming~\cite{Knuth1984}, namely the
idea of \emph{chunks} of knowledge that can be written in a different
order than what the underlying programming language may want, as well as
the idea of tightly coupling explanation with its language encoding.
The weaving of multiple languages together, as available via
org-mode's babel-mode~\cite{schulte2012multi} to achieve ``reproducible
research'' is a more modern take on literate software development.
Jupyter notebooks~\cite{kluyver2016jupyter} is a GUI-driven approach
that is more beginner-friendly but has scaling drawbacks. We have
explicitly decided that it is too early to provide this kind of GUI
for Drasil.

Thinking of doing large scale work has been strongly influenced by Eelco
Visser's grand projects. His work on the Spoofax language workbench~\cite{Spoofax}
made it clear that many different artifacts can be generated. He was braver than
us: his WebDSL~\cite{WebDSL}, at the heart of researchr~\cite{researchr}, is
indeed expected to be long lived, but certainly was not created for a
well-understood domain! The domain analysis of web applications is a non-trivial
contribution of WebDSL, and may well have moved that domain into the ``well understood''
column. Whether it is product lines or program families, weaving
them from DSLs is also something he preached~\cite{voelter2011product}.
We never got a chance to present Drasil to Eelco, but we hope that it would
have resonated with him.

\section{Concluding Remarks}
\label{sec:concluding-remarks}

Dines~\cite{bjorner2021domaineng} has already remarked that for most software,
the \emph{domain knowledge} is the slowest moving part of the requirements.
We add to this the ideas that for certain kinds of software, there is significant
\emph{knowledge duplication} amongst the artifacts, much of which is traceable
to domain knowledge. Thus, in well-understood domains, it should be feasible to
record the domain knowledge ``once'', and then write recipes to generate
instances based on refinements. For long-lived software, this kind of up-front
investment should be worthwhile.

In other words, if we capture the fundamental domain knowledge of specific
domains (such as mechanics of rigid body motion, dynamics of soil, trains,
etc), most later development time can be spent on the \emph{specifics} of the
requirements of a specialized application that may well contain novel ideas at
the refinement or recipe level.

In Drasil, as a side effect of organizing things as we have, we obtain
traceability and consistency, by construction.  Tracking where we use each
concept, i.e. traceability, is illustrated in Figure~\ref{Fig_DrasilAndChange}.
We obtain consistency in our documentation by generating items such as the
table of symbols from those symbols actually used in the document, and whose
definition is automatically extracted from the base knowledge.

There are further ideas that co-exist smoothly with our framework, most notably
software families, software product lines and correctness, particularly
correctness of documentation.  As we generate the documentation in tandem with,
and using the same information as, the source code, these will necessarily be
synchronized. Errors will co-occur in both.  This is a feature, as they are
more likely to be caught that way.

We've also noticed that we can more easily experiment with ``what if'' scenarios,
which make it easy to understand the ramifications of proposed changes.  

We expect to use formal ontologies as we implement
more coherence checks on the domain knowledge itself. For example, it should not
be possible to associate a \textit{weight} attribute to the concept
\textit{program name}, but only to concepts that are somehow ``physical''. 
Perhaps a large ontology in the style of Cyc~\cite{lenat1995cyc} would help.
We also hope to develop a language of design to embody patterns of choices.
Many projects already mentioned have such a language that we are eager to
borrow from.

More people should resurrect older ideas that were too ambitious for their
time, and apply modern understanding and tooling to them.


\bibliography{References}

\end{document}